\documentclass[twocolumn,footinbib,aps,prb]{revtex4-1}
\usepackage{color}
\usepackage{amsfonts,amssymb,amsmath,mathrsfs}
\usepackage{amsbsy}
\usepackage{graphicx}
\usepackage{hyperref}

\newcommand{\bra}[1]{\left\langle #1 \right|}
\newcommand{\ket}[1]{\left| #1 \right\rangle}

\begin{document}

\title{Pulse sequence designed for robust C-phase gates in SiMOS and Si/SiGe double quantum dots}

\author{Utkan G\"ung\"ord\"u}
\email{utkan@umbc.edu}
\affiliation{Department of Physics, University of Maryland Baltimore County, Baltimore, MD 21250, USA}

\author{J.~P.~Kestner}
\affiliation{Department of Physics, University of Maryland Baltimore County, Baltimore, MD 21250, USA}

\begin{abstract}
We theoretically analyze the errors in one- and two-qubit gates in SiMOS and Si/SiGe spin qubit experiments, and present a pulse sequence which can suppress the errors in exchange coupling due to charge noise using ideal local rotations.  In practice, the overall fidelity of the pulse sequence will be limited only by the quality of the single-qubit gates available: the C-phase infidelity comes out to be $\approx 2.5 \times$ the infidelity of the single-qubit operations. Based on experimental data, we model the errors and show that C-phase gate infidelities can be suppressed by two orders in magnitude.

Our pulse sequence is simple and we expect an experimental implementation would be relatively straightforward. We also evaluate the performance of this gate against $1/f$ noise.  Assuming a soft ultraviolet cutoff, we show that the pulse sequence designed for quasistatic noise still performs well when the cutoff occurs below $\sim 1$MHz with experimentally achievable one-qubit Rabi frequencies, suppressing the infidelity by an order of magnitude compared to the existing direct adiabatic protocol. We also analyze the effects of nonadiabaticity during finite rise periods, and find that adiabaticity is not a limitation for the current values of exchange coupling.
\end{abstract}

\maketitle

\section{Introduction}
Silicon is emerging as a viable platform for realizing fault-tolerant quantum computation due to its long coherence times and zero nuclear spin of its most abundant isotope, $^{28}$Si. Spinful isotopes such as $^{29}$Si in natural silicon can be removed down to a concentration of 800ppm \cite*{Yoneda2017b,Chan2018} or lower \cite{Abrosimov2017}. Recently, one-qubit fidelities above the typical 99\%  fault-tolerant fidelity threshold of surface codes \cite*{Fowler2012} have been reported in semiconductor quantum dot spin qubits using isotopically purified $^{28}$Si: 99.9\% in Si/SiGe \cite*{Yoneda2017b} and above 99.9\% in SiMOS \cite*{Veldhorst2014b,Chan2018}. However, two-qubit gate infidelities are two orders of magnitude worse \cite*{Brunner2011,Veldhorst2014,Watson2017,Zajac2017a}. Recent theoretical work predicts that systematic errors can be removed to increase two-qubit fidelities above 97\% \cite*{Russ2017a}, but the fidelity is ultimately limited by charge noise, a fluctuation in the electrostatic dot potential empirically measured to have something like a $1/f$ power spectral density, commonly believed to be caused by charge traps near the interface \cite*{Culcer2009,Jock2018}.

The charge noise can affect one-qubit gate operations through $g$-factor renormalization \cite*{Veldhorst2014} and two-qubit operations through its effect on the exchange interaction via tunneling and ``detuning," i.e., the energy bias between dots.  When the exchange interaction is turned on via biasing to an asymmetric double-well potential \cite*{Brunner2011,Veldhorst2014,Watson2017}, it is typically detuning noise that is dominant.
The sensitivity of the exchange to charge noise can be reduced, but not eliminated, by symmetric operation \cite*{Reed2016a,Martins2016,Zajac2017a}. Suppressing the overall noise in the exchange remains a general and crucial challenge for realization of fault-tolerant two-qubit gates in silicon quantum dots.

In this paper, we address this problem by showing that a robust two-qubit gate can be implemented in existing devices by using a simple pulse sequence which completely removes the leading order effects of the low-frequency exchange noise from the entangling gate. The rest of the paper is organized as follows.
Sections \ref{sec:model} and \ref{sec:adiabatic} contain the background information. In Section \ref{sec:model}, we present the model we use to describe the SiMOS quantum dots, along with the experimental parameters we use. In Section \ref{sec:adiabatic}, we briefly describe the non-robust adiabatic entangling gate used in earlier experiments, and discuss how its fidelity is impacted by diabatic corrections during pulse ramps as well as quasistatic charge noise.  In Section \ref{sec:robust}, we present our main results on realization of a robust perfect entangling gate in SiMOS using the adiabatic gate in conjunction with one-qubit rotations as building blocks. We analyze the robustness of our pulse sequence in the presence of quasistatic as well as time-dependent $1/f$ charge noise. Section \ref{sec:conclusion} concludes the paper.

\section{Model}
\label{sec:model}

The double quantum dot in the $(1,1)$ charge region, with the possibility of tunneling from left to right to the state $(0,2)$, can be modeled using the lab-frame Hamiltonian \cite*{Yang2011,DasSarma2011}
\begin{align}
H = \begin{pmatrix}
E_z & \frac{E_{2,\perp}^*}{2} & \frac{E_{1,\perp}^*}{2} & 0 & 0 \\
\frac{E_{2,\perp}}{2} & \frac{\Delta E_z}{2} & 0 & \frac{E_{1,\perp}^*}{2} & t_0 \\
\frac{E_{1,\perp}}{2} & 0 & -\frac{\Delta E_z}{2} & \frac{E_{2,\perp}^*}{2} & -t_0 \\
0 & \frac{E_{1,\perp}}{2} & \frac{E_{2,\perp}}{2} & - E_z & 0 \\
0 & t_0 & -t_0 & 0 & U-\epsilon
\end{pmatrix}
\end{align}
in the basis of $\ket{\uparrow \uparrow}, \ket{\uparrow \downarrow}, \ket{\downarrow \uparrow}, \ket{\downarrow \downarrow}, \ket{S(0,2)} $. Here $E_z = \mu_B (g_1 B_z^1 + g_2 B_z^2)/2$ is the average Zeeman energy of the electrons in dots due to ``longitudinal" magnetic field, $\Delta E_z = \mu_B (g_1 B_z^1 - g_2 B_z^2) $ is their difference, $t_0$ is tunneling energy, $E_\perp^k$ is the contribution from the ``transverse" magnetic fields $\mu_B g_k (B_x^k + i B_y^k)$ as seen by the $k$th electron, $U$ is the charging energy and $\epsilon$ is the chemical potential which is proportional to the applied gate voltage through lever-arm coefficient $\alpha$. For a single-tone drive, $E_\perp^k$ can be written as $\Omega_k e^{i \omega t}$ where $\Omega_k/h$ is referred to as the one-qubit Rabi frequency and $\omega$ as the microwave frequency.

In the absence of a current through the ESR line ($B_x = B_y =0$), assuming that tunneling is weak ($U-\epsilon \gg t_0$), we use Schrieffer-Wolff transformation \cite*{Schrieffer1966} to block-diagonalize the $(1,1)$ and $(0,2)$ sectors of the Hamiltonian, obtaining
\begin{align}
H_{(1,1)} = \begin{pmatrix}
E_z & 0 & 0 & 0 \\
0 & \frac{\Delta E_z}{2} - \alpha_+ & \frac{\alpha_+ + \alpha_-}{2} & 0 \\
0 & \frac{\alpha_+ + \alpha_-}{2} & -\frac{\Delta E_z}{2} - \alpha_- & 0 \\
0 & 0 & 0 & - E_z
\end{pmatrix}
\label{eq:H}
\end{align}
up to higher order terms in $t_0/(U-\epsilon)$, where $\alpha_\pm = t_0^2/(U-\epsilon \mp \Delta E_z/2)$. Note that $J \equiv \alpha_+ + \alpha_-$ can be identified as the strength of an effective Heisenberg coupling between the two electrons. Using Pauli matrices, this Hamiltonian can also be written in the form
\begin{align}
H_{(1,1)} =& \frac{XX + YY}{2} \frac{J}{2} + \frac{ZI-IZ}{2} \frac{h_z}{2} +\nonumber\\
& ZZ \frac{J}{4} - II \frac{J}{4} + \frac{ZI + IZ}{2} E_z,
\end{align}
with $h_z \equiv \Delta E_z + \alpha_- - \alpha_+$.

We remark that the two generators on the first line above form an $\mathfrak{su}(2)$ subalgebra of $\mathfrak{su}(4)$, and the generators on the second line commute with everything else, forming two $\mathfrak u(1)$ subalgebras which commute with everything else. (The identity term which trivially commutes with everything else is not a Lie generator; in the propogator, it lives in the coset space $\cong \text{U(4)/SU(4)}$, leading to an unimportant global phase factor, therefore, we will drop it in what follows.) It is thus necessary to use transverse magnetic fields, in addition to exchange and longitudinal fields, for building a robust CNOT gate.

We consider the operating regime where $U-\epsilon \gg \Delta E_z$ at all times such that $|\alpha_+ - \alpha_-| \ll \Delta E_z$, which allows us to approximate $h_z \approx \Delta E_z$. Note that  since $g$-factors are electrically modulated, $\Delta E_z$ depends on the applied gate voltage, just as $\epsilon$ and $E_z$ do.

Following the SiMOS experiment \cite*{Veldhorst2014}, we assume that $\epsilon > 0$, and neglect the $\ket{(2,0)}$ state with high energy $U'+\epsilon$. This orbital can be taken into account by a renormalization of the exchange as $J \approx 2 t_0^2[(U'+\epsilon)^{-1}+(U-\epsilon)^{-1}]$ \cite*{Meunier2011}.

For numerical results in what follows, we will use the material parameters from SiMOS quantum dots \cite*{Veldhorst2014} unless specified otherwise: $U=\alpha \times  0.11\text{V} $, $\alpha = \partial \epsilon/\partial V_{G_1} = 0.2$eV/V, $E_z/h =  39.14$GHz, $t_0/h =  900$MHz, $\Delta E_z^{(0)}/h =  39.68$MHz, $\Omega_1/h = 400$MHz, $\Omega_2/h = 360$MHz and $h$ is the Planck constant. We will also consider pulsing $\epsilon$  from $\epsilon^0 = 0$ to $\epsilon^* \approx \alpha \times 102$mV, at which $T_2^*|_{\epsilon=\epsilon^*} \approx 7.15\mu$s \cite*{Veldhorst2014} which approximately corresponds to a 78kHz RMS error in $J/h$ (corresponding to $28 \mu \text{eV}$ RMS error in $ \epsilon$ when all noise is attributed to $\delta \epsilon$). $\Delta E_z$ depends linearly on the gate voltage as  $\Delta E_z^{(0)} + b V_{G_1}$ around $\epsilon=\epsilon^*$:
\begin{align}
\Delta E_z(\epsilon) \approx \Delta E_z^{(0)} + b \epsilon/\alpha,
\end{align}
with $b/h = 2 \times  19\text{MHz/V}$. In this detuning regime, the conditions $U-\epsilon \gg t_0$ and $|\Delta E_z| \gg |\alpha_+ - \alpha_-|$ are well satisfied, since they translate to $400\text{GHz} \gg 900\text{MHz}$ and $40\text{MHz} \gg 0.2\text{kHz}$, respectively.

\section{Adiabatic C-phase gate}
\label{sec:adiabatic}
C-phase is a natural two-qubit gate in the context of single spin-qubits in semiconductor quantum gates. An implementation based on adiabatic evolution within the singlet-triplet subspace has been described in  Ref.~\onlinecite*{Meunier2011}, and this gate was later experimentally realized in SiMOS double quantum dots \cite*{Veldhorst2014}. This implementation, which only involves a simple pulsing of the detuning, is however susceptible to charge noise and care must be taken during pulsing to prevent diabatic transitions.
In this section, we go over the basic idea of the adiabatic gate and show that diabatic transitions can be avoided with a nonlinear ramping profile. The noise will be analyzed in Section \ref{sec:robust}.

\subsection{A non-robust adiabatic C-phase gate}
The adiabatic evolution of the singlet-triplet states of the Hamiltonian Eq.~\eqref{eq:H} can be used to implement a C-phase \cite*{Veldhorst2014} using the adiabatic evolution of the eigenvectors of the Hamiltonian; the nontrivial entangling operation is due to the middle the middle $2\times2$ block of the Hamiltonian Eq.~\eqref{eq:H}, which influences the SU(2) subspace spanned by $\ket{\uparrow \downarrow}$ and $\ket{\downarrow \uparrow}$ states.
We can easily see that such a Hamiltonian can lead to a useful two-qubit gate as follows. The adiabatic theorem guarantees that when the Hamiltonian is varied slowly, the eigenvectors of the Hamiltonian evolve by acquiring a phase without any transitions. In the basis of these adiabatic vectors, a cyclic Hamiltonian results in the unitary time-evolution $U = \text{diag}(1,e^{i \phi_+}, e^{i \phi_-},1)$, which is equivalent to a C-$\pi$-phase gate when $\phi_+ + \phi_- = \pi$, up to local operations.

Specifically, the eigenvectors of the Hamiltonian and their corresponding eigenenergies are given by
\begin{align}
E_\pm =& \frac{1}{2} (-J \pm \Delta E), \quad \Delta E = \sqrt{J^2 + h_z^2},\nonumber\\
\ket{\psi_+} =& \frac{1}{A}\begin{pmatrix}1+\cos\beta \\ \sin\beta\end{pmatrix}, \quad \ket{\psi_-} = \frac{1}{A}\begin{pmatrix}-\sin\beta \\ 1+\cos\beta\end{pmatrix}
\label{eq:eigen}
\end{align}
where $A = \sqrt{2+2\cos\beta}$, $\cos\beta = h_z/\sqrt{J^2 + h_z^2}$, and $\sin\beta = J/\sqrt{J^2 + h_z^2}$. Using the adiabatic theorem, we find that in this subspace, the time-evolution operator is given by
\begin{align}
U_\text{ad}'(t;0) = \sum_{s \in \{+,-\}} e^{-\frac{i}{\hbar} \int_0^t E_s(t') dt'}\ket{\psi_s (t)} \bra{\psi_s (0)}
\end{align}
in the basis of $\ket{\psi_s (0)}$.
Above, the time-evolution operator contains only dynamical phases since the Berry phases $\gamma_s^g = i \oint \bra{\psi_s (t)}d \ket{\psi_s (t)} $ are zero given that the integrand vanishes for real wavefunctions.

\begin{figure}
	\includegraphics[width=1\columnwidth]{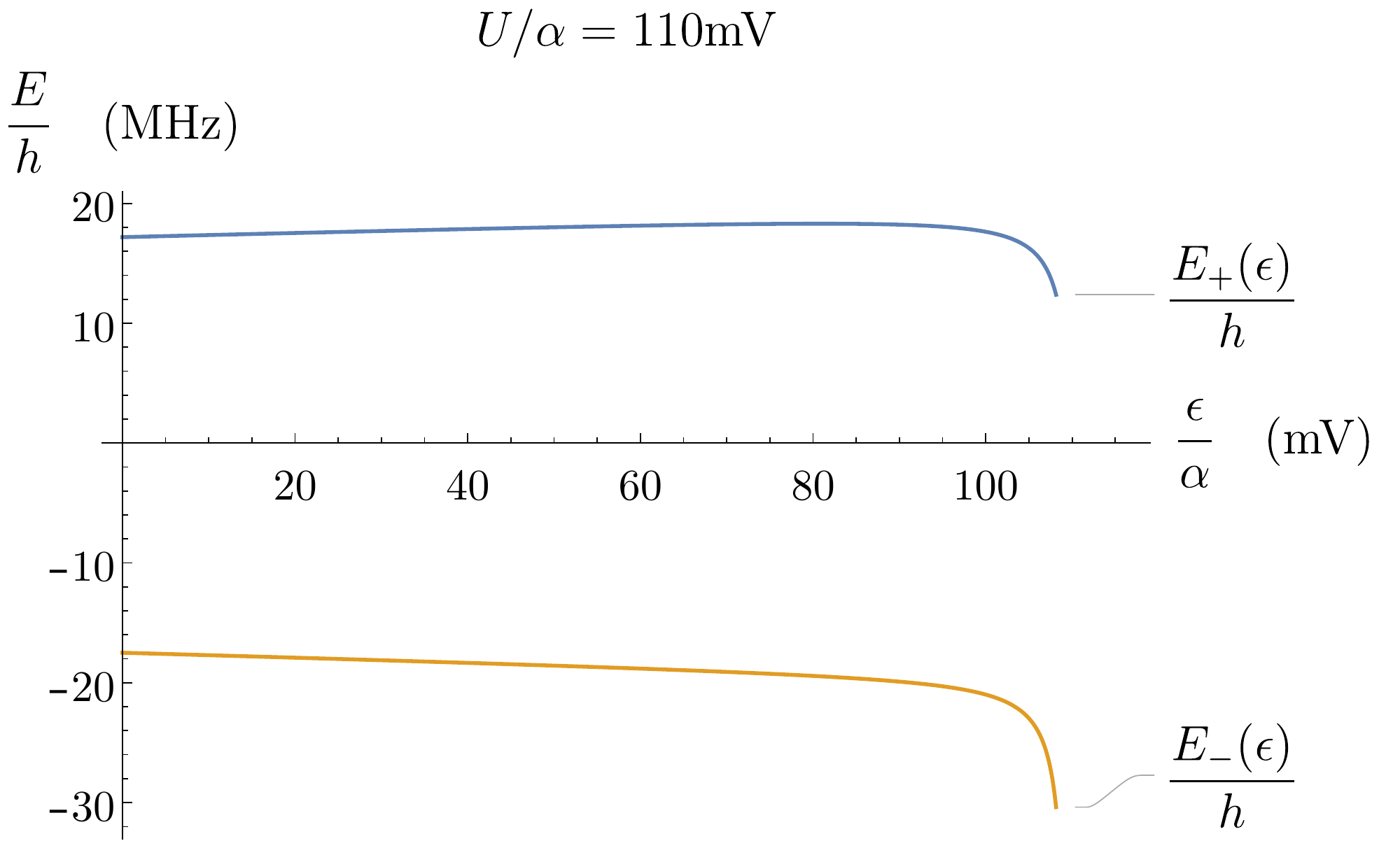}
	\caption{Energy levels of the adiabatic $\ket{\psi_\pm}$ states, as a function of detuning $\epsilon$.}
	\label{fig:energy-levels}
\end{figure}

\subsection{Logical basis for adiabatic quantum computation}
Considering a cyclic evolution in the parameter space, such as ramping up voltage adiabatically and coming back to the initial value, i.e. $\ket{\psi_s (T)} = \ket{\psi_s (0)}$ where $T$ is the desired gate time, the gate operation $U_\text{ad}'(T;0)$ is a diagonal matrix in the basis of $\{\ket{\psi_s (0)}\}$. Note that since the Hamiltonian cannot be turned off completely in the $(1,1)$ charge region, the eigenstates of the Hamiltonian never coincide with spin-eigenstates $\ket{\uparrow \downarrow}$ and $\ket{\downarrow \uparrow}$. For practical purposes, it is preferable to use the eigenvectors of the ``base" Hamiltonian $H_0 = H_{(1,1)}(t=0) = H_{(1,1)}|_{\epsilon=0}$ as the logical basis for quantum computation since these states are stationary when the control parameter $J$ is set to zero \cite*{Ghosh2013,VanDijk2018}. It is this time-dependent basis that has been used as the logical basis for quantum operations in the experiment \cite*{Veldhorst2014}, and we will adapt it as our logical basis in what follows too. In this basis, the adiabatic evolution is given by
\begin{align}
U_\text{ad}(t;0) = \sum_{s \in \{+,-\}} e^{-\frac{i}{\hbar} \int_0^t [E_s(t')-E_s(0)] dt'}\ket{\psi_s (t)} \bra{\psi_s (0)}.
\end{align}
One-qubit operations in this logical basis are nontrivial and are discussed in detail in Appendix~\ref{sec:one-qubit}.

\subsection{Limits of adiabatic control}
\label{sec:adiabatic-limits}
Adiabaticity, while convenient for obtaining an analytical expression for the gate operation, constrains how fast the exchange can be pulsed. This constraint can be quantified in terms of the probability of unwanted transitions due to diabatic terms.
In the basis of time-dependent energy eigenvectors, the middle block of the Hamiltonian including the off-diagonal diabatic terms is
\begin{align}
H_\text{ad} = \begin{pmatrix}
E_+ & V \\
V^* & E_-
\end{pmatrix}
\end{align}
where $V = i \hbar \bra{\psi_\pm(t)}\partial_t\ket{\psi_\mp(t)} = i \hbar \dot \beta$. A loose condition on suppressing diabatic transitions can be obtained by ensuring that the crossing between the adiabatic states is avoided: $|V| \ll |\Delta E|$ or
\begin{align}
\frac{\hbar}{2}\left|\frac{\dot J h_z - J \dot h_z}{J^2 + h_z ^2}\right| \ll \sqrt{J^2 + h_z^2},
\end{align}
at all times during the gate operation. For small $V$ \footnote{We cross-checked that first-order perturbation agrees well with the numerical solution of the Schr{\"o}dinger equation for the set of parameters and pulse shapes mentioned in the text.}, a tighter bound on transition probability can be obtained by using the first-order perturbation as
\begin{align}
P_\text{flip} =& \left|\int_0^{T_\text{ramp}} dt e^{\frac{i}{\hbar} \Delta E t} \frac{V}{\hbar}\right|^2 \nonumber\\
=& \left|\int_0^{T_\text{ramp}} dt e^{\frac{i}{\hbar} \sqrt{J^2 + h_z^2} t} \frac{1}{2}\frac{\dot J h_z - J \dot h_z}{J^2 + h_z ^2} \right|^2 \ll 1.
\end{align}

We observe from the eigenenergies of these adiabatic states in Fig.~\ref{fig:energy-levels} that the energy gap is smallest around the idle gate voltage level $\epsilon^0/\alpha$ and suddenly starts to get large around $\epsilon^*/\alpha$. Therefore, the detuning pulse should be designed such that $\dot{\epsilon}(t)$ gets smaller as $\epsilon$ approaches $\epsilon^*$.
A naive linear ramp from $\epsilon(0) = \epsilon^0$ to $\epsilon(T_\text{ramp}) = \epsilon^*$ is an ill-suited choice and requires a ramp time greater than 200ns to suppress the flip probabilities below an acceptable threshold of $10^{-4}$. With a $\tanh$ pulse \cite*{Motzoi2009} $\epsilon(t) = \epsilon^0 + (1/2)(\epsilon^*-\epsilon^0)[\tanh(t/4T_\text{ramp}) - \tanh([t-T_\text{pulse}]/4T_\text{ramp})]$ on the other hand, we find that the flip probability is still $\sim 10^{-6}$ for a ramp time as short as 10ns.

While it is also possible to suppress the diabatic terms using a pulse sequence \cite*{Russ2017a}, a shaped voltage ramp has the advantage of being faster and simpler.

\section{Robust adiabatic C-phase gate}
\label{sec:robust}
\subsection{Quasistatic noise}
\label{sec:quasistatic}
The adiabatic entangler we described so far is susceptible to charge noise. Nevertheless, we can use it as the building block of a pulse sequence to construct a gate that is equivalent to a C-phase gate up to local unitary operations.

We initially consider random quasistatic (i.e., constant on the timescale of a gate operation) charge noise affecting the detuning, tunneling and $g$-factors, leading to noise in both exchange and one-qubit Rabi frequencies \cite*{Wu2014,Dial2013}.  This affects both one- and two-qubit operations.  We model the noise as a Gaussian distribution, the RMS width of which can be obtained from the $T_2^*$ of a Ramsey experiment \cite*{Schriefl2006},
\begin{align}
T_2 ^* = \left|\frac{\sqrt 2 \hbar}{ \partial_J E_\text{gap} \sigma_{\delta J} }\right|
\label{eq:T2star}
\end{align}
where $E_\text{gap}$ is the energy gap between the two states used for $T_2^*$ measurements and $\sigma_{\delta J}$ is the RMS value of the noise in exchange, due to fluctuations in detuning and tunneling caused by charge noise \cite*{VanDijk2018}.

$T_2^*$ measurements for the exchange in Ref.~\onlinecite*{Veldhorst2014} are realized by turning on the exchange and turning off the transverse fields, and the gap is given by $E_+$. In the experiment corresponding to the converse situation which is used for measuring the $T_2^*$ times (and $\sigma_{\Omega_i}$) for one-qubit operations \cite*{Veldhorst2014b}, the gap is given by $\Omega_i$.

To see the effect of the quasistatic charge noise, we write the unitary time-evolution operator in the logical basis:
\begin{align}
U_\text{ad}(T) =&  e^{-\frac{i}{\hbar} \int_0^T dt \{[J(t) - J(0)]  \frac{ZZ}{4} + \frac{\Delta E(t) - \Delta E(0)}{2} \frac{ZI-IZ}{2}\}} \times \nonumber \\
& e^{-\frac{i}{\hbar} \int_0^T dt \{[E_z(t)  - E_z(0) ]\} \frac{ZI+IZ}{2}}.
\label{eq:Uad}
\end{align}
For a simple square pulse, the nonlocal phase acquired is $[J(\epsilon^*) + \delta J - J(\epsilon^0)] T/4\hbar$ where $\delta J$ is the random quasistatic shift in the exchange. When the nonlocal phase is $\pi/4$, this gate is local-unitarily equivalent to a C-phase gate, accompanied by local $Z$ rotations which can be removed as we discuss below.

The noise in exchange affects both local and nonlocal parts of the adiabatic gate.
However, when the magnetic energy gradient $\Delta E_z$ is much larger than the exchange $J$, which is the regime we focus on here,  the leading error in the $IZ-ZI$ term is $\sim \delta J/\sqrt{J^2 + h_z^2}$, which leads to a negligible error in the order of $\sim \delta J^2$. Thus, the dominant effect of charge noise on the adiabatic gate is only on the nonlocal phase.

The nonlocal part of the evolution can be isolated by applying local $Z$ operations (implemented in software by changing the phase of the microwave drive) to ``unwind" the deterministic $ZI$ and $IZ$ rotations above that naturally accompany the $ZZ$ rotation. Random flip-flops of remnant $^{29}$Si nuclear spins can cause stochastic local $Z$ rotations, but this issue can be dealt with via increased isotopic purification. Alternatively, if the presence of $^{29}$Si nuclei is unavoidable, or similarly, when the effect of the charge noise on electron $g$-factors is not negligible, all local $Z$ rotations can still be echoed out in a robust way as described in Appendix \ref{sec:echo}. We denote the adiabatic time-evolution with $IZ$ rotations canceled as ${\bar U}_\text{ad}(T)$.

At this point, we are left with a noisy nonlocal $ZZ$ rotation, which can be made robust against charge noise up to third order using a 5-step BB1 (BroadBand 1) sequence \cite*{Jones2002}. However, generally speaking, when the Hamiltonian contains an entangling term such as $ZZ$ and local terms, it is possible to implement a significantly shorter robust quantum gate by using known one-qubit robust pulse sequences through an isomorphism which maps one-qubit $\text{SU}(2)$ operations to an $\text{SU}(2) \subset \text{SU}(4)$ containing $ZZ$ rotations and two distinct local rotations \cite*{Ichikawa2013}. We will make use of this latter route,  using the 3-step minimal entangling sequence described in Ref.~\onlinecite*{Ichikawa2013} that is robust up to second order against a fixed $J$-coupling error, which corresponds to a generalized version of {\sc scrofulous}  (Short Composite ROtation For Undoing Length Over-- and UnderShoot \cite*{Cummins2003}) under the mentioned mapping, and is given by
\begin{align}
U_\text{seq} = e^{-i \zeta ZZ} e^{i \frac{\theta}{2} IX} e^{-i \frac{\pi}{2} ZZ} e^{-i \frac{\theta}{2} IX}  e^{-i \zeta ZZ},
\label{eq:SCROFULOUS}
\end{align}
where the rotation angles are \footnote{Under the constraint that \unexpanded{$\zeta > 0$} \cite*{Ichikawa2013}, the inverse of the sinc function at \unexpanded{$\sqrt {2/\pi}$} is unique.}
\begin{align}
\zeta = -\frac{\pi}{4} \sec\theta, \quad
 \sec\theta = \frac{2}{\pi} \text{sinc}^{-1}\frac{\sqrt 2}{\pi} \approx -1.280.
 \label{eq:SCROFULOUS-angles}
\end{align}
Furthermore, from the Cartan decomposition of $U_\text{seq}$,
\begin{align}
U_\text{seq} = e^{-i \frac{\eta}{2} IX} e^{-i \frac{\pi}{4} ZZ} e^{i \frac{\eta}{2} IX}
\end{align}
where $\tan\eta = \tan\theta \sec\left( \frac{\pi}{2} \sec\theta \right)$, we see that this gate is local-unitarily equivalent to a C-phase gate.
Thus, the pulse sequence $U_\text{seq}$ is an entangling gate that is robust against the noise in exchange. (For a generalization to different values of $\delta J$ at different times in the three entangling stages, which may be required in setups with bandwidth constraints, see Appendix~\ref{sec:bandwidth}.)

Hence, in terms of $\bar U_\text{ad}$, the overall pulse sequence is
\begin{align}
U_\text{seq} =& {\bar U}_\text{ad}\left(\frac{\hbar \zeta }{J_\text{eff}/4}\right) e^{i \frac{\theta}{2} IX} {\bar U}_\text{ad}\left(\frac{\hbar \pi/2 }{J_\text{eff}/4}\right) \times \nonumber\\
& e^{-i \frac{\theta}{2} IX}  {\bar U}_\text{ad}\left(\frac{\hbar \zeta }{J_\text{eff}/4}\right)
\end{align}
where $J_\text{eff} = J(\epsilon^*) - J(\epsilon^0)$. This assumes a simple square pulse. Since the Hamiltonian required for each segment of the pulse sequence commutes with itself at different times, finite ramping times for detuning and one-qubit Rabi frequencies can be handled exactly, resulting, e.g., in slightly larger time values when using a shaped ramp such as the tanh ramp described earlier (see Appendix~\ref{sec:ramp}).

Compared to the BB1-based pulse sequence\cite*{Jones2002} (which can suppress quasistatic errors in gate operation up to third order), this pulse sequence has about half as many entangling operations and so runs about twice as fast for a CNOT gate when assuming arbitrarily fast one-qubit gates. In the experimentally realistic situation with slow one-qubit ESR gates, the benefits are even more pronounced due to the fewer one-qubit gates required, resulting in a CNOT gate about seven times faster than the BB1 sequence. In general, to realize a nontrivial robust $ZZ$ rotation by an angle $\xi$, the minimal sequence takes $\hbar [(2\zeta+\pi/2)/J_\text{eff} + 2\theta/\Omega]$ time in total whereas the BB1-based sequence takes $\hbar [5\arccos(-\xi/4\pi)/\Omega + 2(4\pi+\xi)/J_\text{eff}]$.

We now quantify the robustness of the pulse sequence using the state-averaged gate fidelity (which is integrated over the H\"aar measure) between the ideal evolution $U$ and the noisy evolution $\tilde{U}$,
\begin{align}
\mathcal F = \frac{1}{N} + \frac{1}{(N+1)N^2} \sum_{i=1}^{N^2-1} \text{tr}\left( \tilde U \Lambda_i \tilde U^\dagger U \Lambda_i U^\dagger \right)
\end{align}
where $\Lambda_i$ denotes SU($N=4$) generators $\sigma_a \otimes \sigma_b$ for two-qubit gates and SU($N=2$) generators $\sigma_i$ for one-qubit gates \cite*{Cabrera2007a}. The noise-averaged infidelity of the pulse sequence is
\begin{align}
\langle 1- \mathcal F_\text{seq}\rangle =& \left\langle \frac{\pi^4 \tan^2\theta}{80} \left(\frac{\delta J}{J_\text{eff}}\right)^4 + \mathcal O \left(\frac{\delta J}{J_\text{eff}}\right)^6 \right\rangle \nonumber\\
& \approx 0.78 \times 3 \left(\frac{\sigma_{\delta J}}{J_\text{eff}}\right)^4
\end{align}
to the leading order, where $\langle \ldots \rangle$ denotes averaging over different realizations of the (Gaussian) random noise. Compared to the infidelity of a direct implementation of C-phase using $\bar U_\text{ad}$, which is $(4/5)(\pi/4)^2(\sigma_{\delta J}/J_\text{eff})^2$, the robust pulse sequence diminishes the average infidelity by a factor of $(3/4)(\pi^2 \tan^2\theta)(\sigma_{\delta J}/J_\text{eff})^2 \approx 4.72(\sigma_{\delta J}/J_\text{eff})^2$.

Assuming quasistatic noise and using the $T_2^*$ value at $\epsilon = \epsilon^*$ in Eq.~\eqref{eq:T2star}, we obtain the relative exchange error  $\sigma_{\delta J}/J_\text{eff} = 78\text{kHz}/3.125\text{MHz} \approx 0.025$ using SiMOS parameters.
Similarly, with paramaters from the experiment in Si/SiGe \cite*{Watson2017}, $\sigma_{\delta J}/J_\text{eff} = (11\mu\text{eV} \times 10^{-4})/(6\text{MHz} \times h) \approx 0.044$. In both cases, we find an improvement of two orders in magnitude in infidelity, from $\sim 10^{-3}$  to $\sim 10^{-5}$, when compared to the naive implementation.

We note that a similar estimate of $\sim 10^{-3}$ for the infidelity of the direct implementation due to charge noise has been reported in Ref.~\onlinecite*{VanDijk2018} in Si/SiGe. Randomized benchmarking \cite*{Huang2018} or Bell state tomography experiments \cite*{Watson2017} indicate worse fidelities for composite entangling operations, as these results include other sources of errors which are not compensated for, including contributions from noisy one-qubit gates, noisy idle gates, nuclear spin flips, crosstalk, timing errors, Bloch-Siegert shift, and other systematic errors \cite*{VanDijk2018,Huang2018}.

This has so far assumed that the local $X$ rotations prescribed by the pulse sequence can be implemented robustly.  When this pulse sequence is implemented using non-robust one-qubit gates in the presence of a generic quasistatic noise $e^{-i [(\Omega_2 + \delta\Omega_2) IX + \delta \Omega_2^y IY + \delta \Omega_2^z IZ]  t/2\hbar}$ where $\delta \Omega_2$, the average infidelity of the entangling gate comes out to be
\begin{align}
\langle 1-\mathcal F_\text{seq}\rangle \approx & \int d \delta\Omega_2 d \delta\Omega_2^y d \delta\Omega_2^z p(\delta\Omega_2, \delta\Omega_2^y, \delta\Omega_2^z) \times \nonumber\\
& \frac{4}{5}\left[ \left(\frac{\delta \Omega_2}{\Omega_2}\right)^2 \theta^2 + \frac{(\delta\Omega_2^y)^2 + (\delta\Omega_2^z)^2}{\Omega_2^2} \sin^2\theta \right],
\label{eq:fidelity-general}
\end{align}
where $p$ is the joint probability density of the noise in a general form, to the leading order in small noise perturbations. Since magnetic noise entering through hyperfine interaction can be remedied by using silicon with a lower concentration of $^{29}$Si, we focus on estimating an upper bound for the electrical noise, which can be due to electrical noise affecting the Rabi frequency via a shift in the valley splitting \cite*{Veldhorst2014b}. As an example, in the particular case of $IX$ noise,
 this simplifies to
\begin{align}
\langle 1 - \mathcal F_\text{seq}\rangle \approx &  \int d \delta\Omega_2 p(\delta\Omega_2)  \frac{ 4}{5} \theta^2 \left(\frac{\delta \Omega_2}{\Omega_2}\right)^2 \nonumber \\
 \approx & 4.87 \left(\frac{\sigma_{\delta \Omega_2}}{\Omega_2}\right)^2,
\label{eq:fidelity}
\end{align}
where we assumed a Gaussian distribution for $\delta \Omega_2$ with a RMS value of $\sigma_{\delta \Omega_2}$. This result is comparable to the average infidelity of a one-qubit $\pi$-pulse
\begin{align}
\langle 1-\mathcal F_{IX_\pi}\rangle = \frac{4}{5} \left(\frac{\pi}{2}\right)^2 \left(\frac{\sigma_{\delta \Omega_2}}{\Omega_2}\right)^2 \approx 1.97 \left(\frac{\sigma_{\delta \Omega_2}}{\Omega_2}\right)^2.
\end{align}
For SiMOS, using the $T_2^*$ value $120\mu$s \cite*{Veldhorst2014,Veldhorst2014b} and $E_\text{gap}=\Omega_2$, we estimate $\sigma_{\delta \Omega_2} = |\sqrt 2 \hbar / \partial_{\Omega_2} E_\text{gap} T_2^*| \approx h \times 1.9\text{kHz}$ yielding $\sigma_{\delta \Omega_2} / \Omega_2 \approx 0.005$.
A similar result can be obtained when $IX$, $IY$ and $IZ$ terms are retained in Eq.~\ref{eq:fidelity-general} as fully correlated charge noise and non-correlated magnetic noise terms.

\subsection{Time-dependent $1/f$ noise}
\label{sec:time-dependent}
While the pulse sequence we have described is effective against quasistatic charge noise which changes at a rate much slower than the pulse sequence, silicon quantum dots suffer from fast noise as well. Noise with $\sim 1/f^\alpha$ power spectral density (PSD) affects a wide range of solid state systems \cite*{Kogan1996}, and is present in silicon quantum dots with $\alpha \approx 1$ \cite*{Yoneda2017b}. We thus analyze the effectiveness of our C-phase gate in the presence of $1/f$ charge noise.

Charge noise introduces electrical fluctuations which affect the exchange $J$ as well as the effective $g$-factors of the spins, which in turn affect their Rabi frequencies $\Omega_i$. We denote the noise Hamiltonian as $H_\varepsilon = \sum_i \chi_i(t)\beta_{i}(t) \Lambda_i$ where $\beta_i(t)$ is the stochastic noise, $\chi_i(t)$ is a dimensionless factor depending purely on the control Hamiltonian at that time that systematically modulates the noise strength, and $\Lambda_i$ is an SU(4) generator. At the operational points we use, the amplitude of these errors are much smaller than the overall strength of the Hamiltonian. This allows us to use a perturbative approach to calculate the influence of the noise.

A practical way of investigating the frequency-dependent robustness of a pulse sequence is the filter function \cite*{Cywinski2008}, which is a measure of susceptibility of the fidelity of a quantum time evolution in response to a noise PSD:
\begin{align}
\langle 1 - \mathcal F_{\text{tr}} \rangle \approx \sum_{i,j=1}^{N^2-1} \frac{1}{\hbar^2} \int_{-\infty}^{\infty} \frac{d\omega}{2\pi} S_{ij}(\omega) \frac{F_{ij}(\omega)}{\omega^2}
\label{eq:gate-infidelity}
\end{align}
for weak noise \cite*{Green2013}, where $\mathcal F_{\text{tr}}$ here denotes the trace fidelity, $|\text{tr}(U \tilde U^\dagger)/\text{tr}(U U^\dagger)|^2$ \footnote{Here, the trace fidelity is preferred for its relatively simple relation to the filter function. While it does not directly compare to the state-averaged fidelity used in earlier section, it can be used to compare the gate fidelities of the pulse sequence and the primitive pulse.}, $S_{ij}(\omega)$ is the PSD of the noise due to the correlation between different stochastic noise components at different times through the two-point correlator
\begin{align}
C_{ij}(|t'-t|) = \langle \beta_{i}(t) \beta_{j}(t')\rangle = \int_{-\infty}^{\infty} d \omega S_{ij}(\omega) e^{i \omega (t'-t)},
\end{align}
and $F_{ij}(\omega)$ is the filter function of the pulse sequence.

To evaluate the filter function, we first write the noise-free time-evolution operator of the pulse sequence $U_\text{seq}(t)$
as a function of time and its adjoint representation as
\begin{align}
R_{ij}(t) = \frac{\text{tr}(\Lambda_i U^\dagger_{\text{seq}}(t) \Lambda_j U_{\text{seq}}(t))}{\text{tr}(\Lambda_i \Lambda_i)}.
\end{align}
Then, to the leading order in noise amplitudes, the filter function is given by \cite*{Green2013,Kabytayev2014} (see Appendix~\ref{sec:filter-function} for details)
\begin{align}
F^{(1)}_{ij}(\omega) = R_{kj}(\omega) R_{ki}^*(\omega)
\end{align}
where
\begin{align}
R_{ki}(\omega) = -i \omega \int_0^{T_\text{seq}} \chi_i(t) R_{ki}(t)  e^{i\omega t} dt.
\label{eq:R}
\end{align}
Here, we moved $\chi_i(t)$ from the PSD into the definition of $R_{ik}(\omega)$, and consequently into the definition of the filter function, such that all terms which depend on the control are collected within the filter function and the remaining stochastic factors can be treated as an effective PSD.

For an SU(4) pulse sequence, the filter-function is a $15 \times 15$ matrix. However, as discussed above, the most significant noise channels present during the pulse sequence are $IX$ and $ZZ$, making the $R_{ik}(\omega)$ matrices very sparse. We further assume that the noise in $J$ and $\Omega_i$ are both mainly due to charge noise, and for simplicity assume that they are fully correlated. Furthermore, we assume that $\chi_i(t)$ does not affect $\beta_i(t)$, or more concretely, that the charge noise (and the lever-arm value), which affects the local spatially averaged scalar potential $\phi$, does not vary with the gate voltage \cite*{Reed2016a} or the current through the ESR line. Under these assumptions, we can write $S_{ij}(\omega) = S_\phi(\omega)$, and the fidelity can be written as
\begin{align}\label{eq:filterfcn-infidelity}
\langle 1 - \mathcal F_{\text{tr}} \rangle \approx \frac{1}{ \hbar^2} \int_{-\infty}^{\infty} \frac{d\omega}{2\pi} S_\phi(\omega) \frac{F^{(1)}(\omega)}{\omega^2}.
\end{align}
The fact that noise is present only for the error channels $i,j \in \{IX,ZZ\}$ is encoded in the filter function through $\chi_i(t)$ which vanishes for all other channels.

The leading order filter function for the pulse sequence is given by
\begin{align}
F^{(1)}(\omega) =  |R_{J}(\omega) + R_{\Omega_2}(\omega)|^2
\end{align}
where
\begin{align}
R_{J}(\omega) =&  \kappa_{J} \sum_{n=1,3,5}\begin{pmatrix}
 (e^{i\omega T_{n}}-e^{i\omega T_{n-1}})  \sin\left(\frac{n-1}{2}\theta\right) \sin (2\zeta) \\
 (e^{i\omega T_{n}}-e^{i\omega T_{n-1}})  \sin\left(\frac{n-1}{2}\theta\right) \cos (2\zeta) \\
-(e^{i\omega T_{n}}-e^{i\omega T_{n-1}}) \cos\left(\frac{n-1}{2}\theta\right) \\
\end{pmatrix}, \nonumber\\
R_{\Omega_2}(\omega) =& \kappa_{\Omega_2} \sum_{n=2,4}\begin{pmatrix}
(e^{i\omega T_{n}}-e^{i\omega T_{n-1}})\cos2\zeta \\
-(e^{i\omega T_{n}}-e^{i\omega T_{n-1}})\sin2\zeta\\
0
\end{pmatrix},
\end{align}
assuming a piecewise-constant control with $\kappa_J = \partial_{\phi}J/4$, $\kappa_{\Omega_2} = \partial_{\phi} \Omega_2/2$, and the vector space on which $R_{J}(\omega)$ and $R_{\Omega_2}(\omega)$ are written above corresponds to the $(IX,ZY,ZZ)$ error channels. $T_n$ denotes the time spent until the $n$th step of the pulse sequence is completed: $T_n = \Theta(5-n) T_\zeta + \Theta(4-n) T_{\theta/2} + \Theta(3-n) T_{\pi/2} + \Theta(2-n) T_{\theta/2} + \Theta(1-n) T_\zeta $ where $T_\zeta = 4\zeta \hbar/J_\text{eff}$, $T_{\theta/2} = (\theta/2) \hbar /\Omega_2$, $T_{\pi/2} = 2\pi \hbar/J_\text{eff}$, and $\Theta(x)$ is the Heaviside step function.

\begin{figure}
	\includegraphics[width=1\columnwidth]{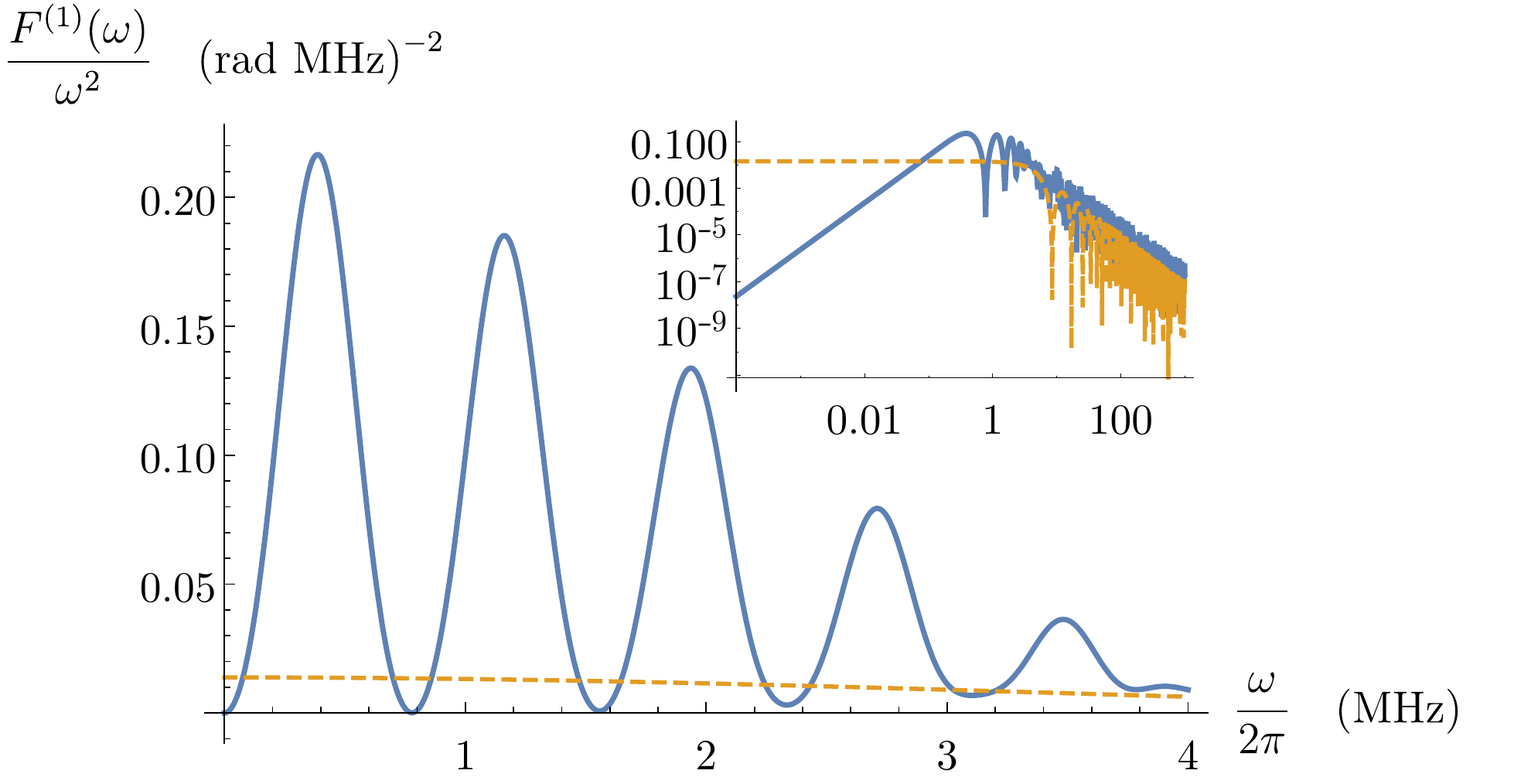}
	\includegraphics[width=1\columnwidth]{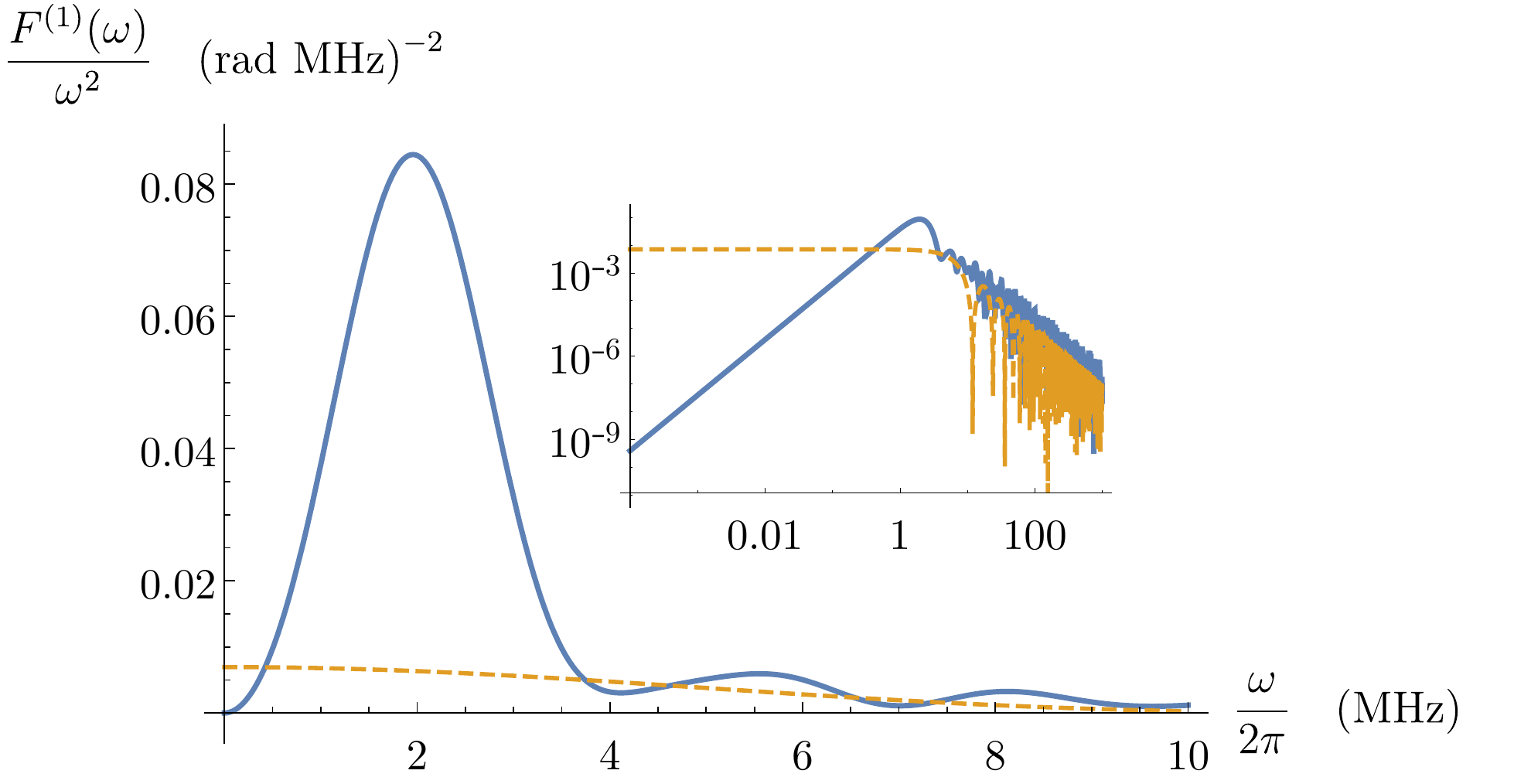}
	\caption{(Color online) Comparison of filter functions for the composite pulse sequence (blue) and a primitive $ZZ_{\pi/4}$-pulse (yellow, dashed) in the presence of exchange noise, with $\delta \Omega_2=0$. The top figure corresponds to SiMOS parameters with $J_\text{eff}/h \approx 4$MHz, $\Omega/h=360$kHz \cite*{Veldhorst2014}, and the bottom figure is using Si/SiGe parameters $J_\text{eff}/h = 6$MHz, $\Omega/h=4$MHz \cite*{Watson2017}. The pulse sequence filters out a significant portion of the noise at low frequencies.}
	\label{fig:filter-function}
\end{figure}

\begin{figure}
	\includegraphics[width=0.8\columnwidth]{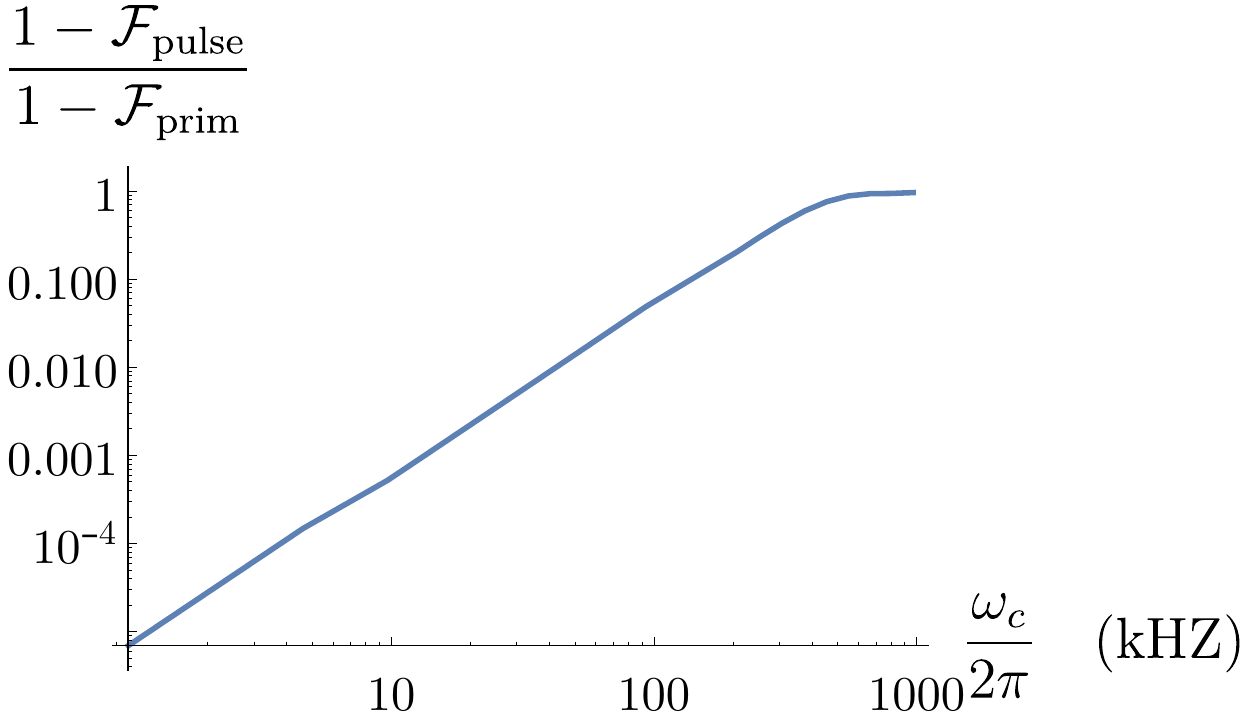}
	\includegraphics[width=0.8\columnwidth]{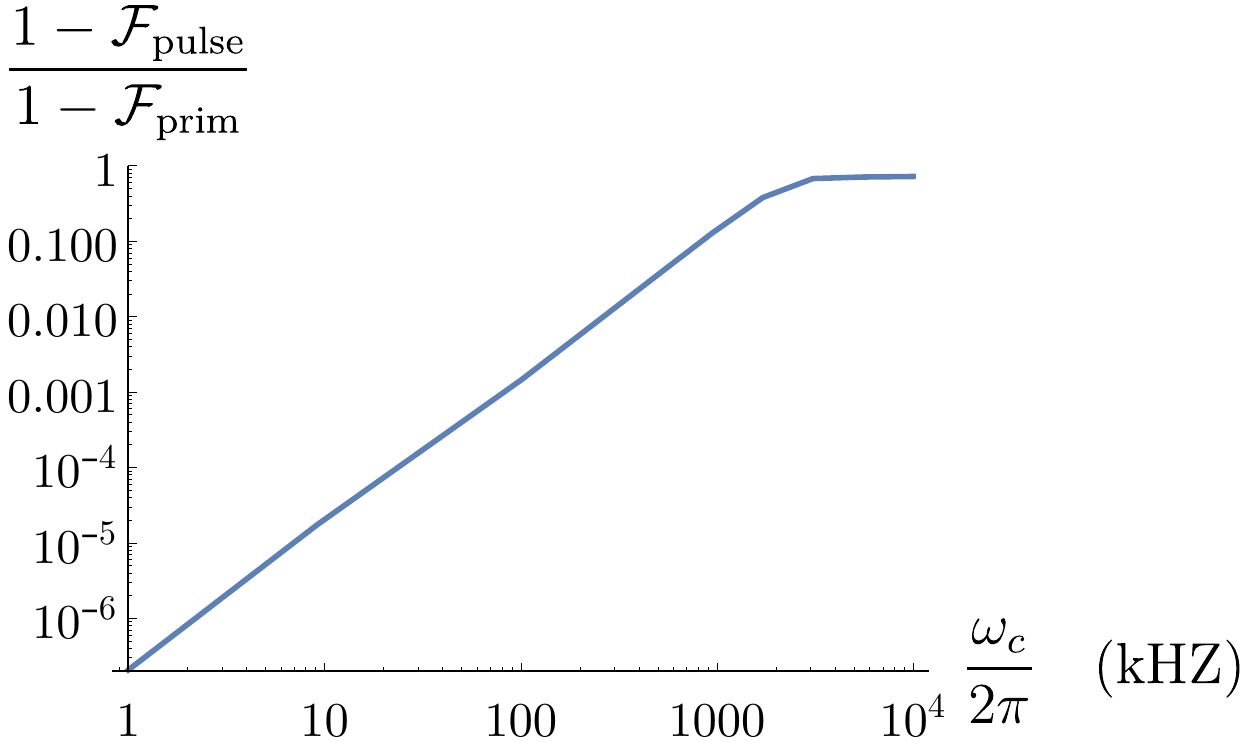}
	\caption{Relative infidelity of the pulse sequence with respect to the infidelity of the primitive pulse, assuming a PSD of the form given in Eq.~\eqref{eq:PSD}, as a function of ultraviolet cutoff frequency, $\omega_c$ for SiMOS (top) and Si/SiGe (bottom) using values given in the caption of Fig.~\ref{fig:filter-function}. The infrared cutoff value is taken to be $\omega_\text{ir}=2\pi/100$s, corresponding to a typical calibration time \cite*{Yoneda2017b}.}
	\label{fig:relative-fidelity}
\end{figure}

When the noise in $\Omega_2$ is negligible compared to noise in exchange $J$ (which is true for our parameters) or when one-qubit $IX$ rotations robust against first order quasistatic noise are used, the low-frequency behavior of the filter function is given by
\begin{align}
F^{(1)}(\omega) = \kappa_J^2 \left[T_\zeta (T_{\pi/2} + T_{\theta/2} + T_\zeta )\sin\theta \right]^2 \omega^4 + \mathcal O (\omega^6).
\end{align}
The fact that the usual lowest order term $\sim\omega^2$ is not present is due to the robustness of the pulse sequence against quasistatic noise.
We have neglected the effects of noise in $\Omega_2$ because, although it gives rise to a contribution of $(\kappa_{\Omega_2} T_{\theta/2} \omega)^2$ to the filter function at low frequencies, for our parameters and taking $\partial_{V} \Omega \approx 29\text{MHz} h/\text{V}$ \cite*{Veldhorst2014b} and $\partial_{\epsilon/\alpha} J|_{\epsilon=\epsilon^*} \approx 541\text{MHz} h/\text{V}$, $\kappa_{\Omega_2}/\kappa_{J} \approx 0.1$, which leads to a small correction. This remains true when similar $IY$ and $IZ$ charge noise terms are included. The full frequency dependence of the filter function (divided by $\omega^2$ for clarity and because this is the quantity that appears in Eq.~\eqref{eq:filterfcn-infidelity}) is shown in Fig.~\ref{fig:filter-function}.

To translate this filter function into an estimated fidelity, we assume a $1/f$ PSD for charge noise with a $1/f^2$ tail: \cite*{Schriefl2006,Cywinski2008,Kabytayev2014}
\begin{align}
S_\phi(\omega) \approx
\begin{cases}
0 & 0 < |\omega| < \omega_\text{ir} \\
       \frac{2\pi A_f}{|\omega|} & \omega_\text{ir} < |\omega| < \omega_c \\
       \frac{2\pi A_f \omega_c}{\omega^2} & \omega_\text{c} < |\omega| \\
\end{cases}
    \label{eq:PSD}
\end{align}
where $\omega_\text{ir}$ and $\omega_c$ denote the infrared and ultraviolet cutoff frequencies of the noise spectrum, and $\sqrt{A_f}$ is the charge noise at 1Hz. Errors that change at a rate slower than inverse experiment time can be calibrated away at the beginning of the experiment, which sets the infrared cutoff value \cite*{Yoneda2017b}; this implies that longer running experiments have smaller $\omega_\text{ir}$. We remark that this PSD approximates a weighted sum of Lorentzian fluctuators over a finite range of characteristic frequencies $\gamma$, that is $\propto \int_{\omega_\text{ir}}^{\omega_c} \frac{1}{\pi \gamma} \frac{\gamma}{\gamma^2+\omega^2} d\gamma$.

From Fig.~\ref{fig:filter-function}, we observe that the pulse sequence filters out low frequency quasistatic noise. This also removes the necessity of frequent recalibrations. However, we also observe from the insets that after a cross-over point, the pulse sequence starts to amplify the noise at higher frequencies. Although the PSD also decays with increasing frequency, a $1/\omega$ decay is typically not fast enough, which makes the ultraviolet cutoff value very important for the design of pulse sequences in general. This is very relevant in our context, because in a recent experiment in Si/SiGe with isotopically purified $^{28}$Si, $1/\omega$ behavior has been reported at least up to 320kHz \cite*{Yoneda2017b}, which is close to the cross-over frequencies shown in Fig.~\ref{fig:filter-function}. In a similar experiment with natural silicon, a possible crossover to $1/\omega^2$ behavior is observed around 200kHz-500kHz \cite*{Kawakami2016}, although the data is inconclusive, and the PSD may indeed vary to a considerable extent between devices. It may also be possible that a cross-over may not exist in the relevant frequency regime for dynamical decoupling schemes.

Using this PSD with an optimistic infrared cutoff value corresponding to 100s without a recalibration, we show how the pulse sequence fares against a primitive $ZZ$ rotation to implement a CNOT gate for a range of ultraviolet cutoffs, as shown in Fig.~\ref{fig:relative-fidelity}. In current experiments, the Rabi frequencies are of limited strength and are significantly lower than the exchange. The slow local gates increase the total duration of our pulse sequence significantly, and our estimates show that it strongly impacts fidelity of the gates in the presence of $1/f$ noise. Using SiMOS parameters with $\Omega/h=360$kHz, we observe that using the pulse sequence improves the infidelity by an order of magnitude for a UV cutoff value $\approx 150$kHz, and remains beneficial up to $\approx 500$kHz.
On the other hand, when using Si/SiGe parameters with a similar exchange value but $\Omega/h=4$MHz, we observe an order in magnitude improvement in infidelity at $\omega_c/2\pi \approx 1$MHz and the pulse sequence remains beneficial in general but the benefits saturate at around $\omega_c/2\pi \approx 3$MHz.

\section{Conclusion}
\label{sec:conclusion}
In this work, we have detailed how a minimal robust pulse sequence for a gate that is local-unitarily equivalent to C-phase can be implemented in a silicon spin qubit system using adiabatic evolution and local rotations. We showed that adiabaticity is not a concern in implementing fast gates in these devices currently, and that gate times are mainly restricted by the strength of exchange and one-qubit Rabi frequencies.  We analyzed the fidelity of our two-qubit pulse sequence in the presence of both quasistatic and time-dependent $1/f^\alpha$ noise by analytically deriving the two-qubit filter function.
For quasistatic noise, the pulse sequence suppresses the infidelities by two orders in magnitude when robust one-qubit gates are used, and causes the two-qubit gate to have essentially the same performance as a one-qubit gate otherwise. For time-dependent $1/f$ noise with a soft UV cutoff crossover to $1/f^2$ and using relevant experimental parameters, we have found that the pulse sequence remains beneficial when the cutoff frequency is below 3MHz (500kHz) for a Rabi frequency of $\Omega/h = 4$MHz\cite*{Watson2017} ($360$kHz\cite*{Veldhorst2014}). This highlights the importance of the cutoff frequency for robust quantum control in Si spin qubits. Although the cutoff frequency has not yet been measured, if it is smaller than the attainable Rabi frequency and one uses a robust one-qubit gate protocol, the pulse sequence we have presented makes it possible to implement a C-phase gate with a fidelity well above the fault-tolerance threshold of surface codes.

\begin{acknowledgements}
We thank Andrew S. Dzurak, C. Henry Yang, Wister Huang and Arne Laucht for helpful discussions. This research was sponsored by the Army Research Office (ARO), and was accomplished under Grant Number W911NF-17-1-0287.
\end{acknowledgements}

\appendix
\section{Smooth pulse ramps}
\label{sec:ramp}
The pulse sequence described in the main text assumes the controllable parameters such as $J$ and $\Omega_i$ can be turned on and off abruptly. This, however, is not actually a requirement since one can operate the qubits such that the Hamiltonian of the system commutes with itself during the ramp periods, e.g., by ensuring that the exchange coupling and the ESR line are never both on at the same time in our proposal.  In that case only the total area of the pulse shape matters for $ZZ$ rotations.

A shaped ramp would be chosen based on bandwidth constraints of the control. In the frequency domain, a tanh pulse (described in Section \ref{sec:adiabatic-limits}), Gaussian \cite*{Jones2018} or an erf pulse \cite*{Ghosh2013} are each well-localized at low-frequencies and either is a suitable choice. The finite ramp time contributes to the overall time-evolution operator, but this can be easily calculated when the Hamiltonian commutes with itself at different times during the ramping. For instance, when pulsing the exchange, the time-evolution operator is given by Eq.~\eqref{eq:Uad} which holds regardless of the time-profile of exchange.

Similar is true for local $X$ rotations; local $Z$ rotations can simply be absorbed into the definition of the logical basis since their Zeeman energies do not vary with applied gate voltages. We stress that the corresponding filter function will depend on the shape of the ramp via the integral given in Eq.~(\ref{eq:R}).

\section{Bandwidth-limited $J$}
\label{sec:bandwidth}
When the exchange cannot be changed quickly during a pulse sequence due to bandwidth limitations, implementing a square pulse becomes impossible. When using a shaped pulse, the value of $J_\text{eff}$ varies in time. Since the electrical sensitivity of the exchange $\partial_V J$ will also vary in time, the average exchange error for each $ZZ$ rotation will be different for different pulse shapes in general.

Due to the symmetry of the pulse sequence, the same pulse shape can be used for the first and last $ZZ$ rotations. However, the rotation angle of the middle $ZZ$ rotation is in general different from the first and last, which means the average exchange error for it will be different from that of the outer $ZZ$ rotations.
This in turn means that the pulse sequence given by Eqs.~\eqref{eq:SCROFULOUS} and \eqref{eq:SCROFULOUS-angles} cannot readily be used in such a situation since it was derived under the assumption that the average exchange error is same for all $ZZ$ rotations.  In this appendix, we give a generalized version of the pulse sequence which allows using different average values of exchange for the outer and middle $ZZ$ rotations, which we will label $J_\text{out}^\text{eff}$ and $J_\text{mid}^\text{eff}$, respectively. A symmetric (as far as $ZZ$ rotations are concerned) pulse sequence with two different exchange levels is sufficient, because if the first half can be implemented, so can the second half.

To realize a robust unitary which corresponds to a $\xi$ rotation around $ZZ$ up to local $IX$ rotations such that $U_\text{target} = \exp(-i \eta IX) \exp(-i \xi ZZ) \exp(i \eta IX)$, the following minimal pulse sequence can be performed:
\begin{align}
U_\text{seq} = e^{-i \zeta ZZ} e^{i \frac{\theta}{2} IX} e^{-i \frac{\pi}{2} ZZ} e^{-i \frac{\theta}{2} IX}  e^{-i \zeta ZZ},
\end{align}
where the one-qubit rotation angle $\theta$ is determined by the target $ZZ$ rotation angle $\xi$ as the numerical solution of
\begin{align}
\cos\xi = \cos\theta\sin\left(c \frac{\pi}{2} \sec\theta \right),
\end{align}
under the constraint that $\sec\theta < 0$ (to ensure that $\zeta>0$, which is given below in Eq.~\eqref{eq:aux-zz}),  with $c =[\delta J_\text{mid}/J_\text{mid}^\text{eff}] / [\delta J_\text{out}/J_\text{out}^\text{eff}]$, where $\delta J_i = \int_{{t_0}_i}^{{t_0}_i + T_i} dt \partial_V J/T_i$ denotes the average electrical sensitivity of the exchange, and similarly, $J_i^\text{eff}$ denotes the average effective exchange $\int_{{t_0}_i}^{{t_0}_i + T_i} dt J_\text{eff}/T_i$, for each $ZZ$ rotation. The value of $c$ can be approximated as $\sim J_\text{mid}/J_\text{out}$ at high enough values of exchange ($J_\text{mid},J_\text{out} \gg J_0$ where $J_0$ is the residual exchange) when detuning noise dominates, and $\sim 1$ when tunneling noise dominates; in the absence of detailed knowledge about the nature of noise, it can be experimentally calibrated by measuring $\delta J$ as a function of $J$. Finally, $\theta$ in turn determines the angle of the auxiliary $ZZ$ rotations as
\begin{align}
\zeta = -c \frac{\pi}{4} \sec\theta.
\label{eq:aux-zz}
\end{align}
In terms of $\theta$, the one-qubit rotations which accompany the $ZZ$ rotations in $U_\text{target}$ are given by
\begin{align}
\tan\eta = \tan\theta \sec\left( c\frac{\pi}{2} \sec\theta \right).
\end{align}
The case $J_\text{mid}^\text{eff} = J_\text{out}^\text{eff}$ corresponds to $c=1$, and with $\xi=\pi/4$ (that is, targeting a CNOT gate), we recover Eq.~\eqref{eq:SCROFULOUS-angles}.

\section{One-qubit local $X$ rotations in the logical frame}
\label{sec:one-qubit}
In the presence of $\Omega_i$, the Hamiltonian after the Schrieffer-Wolff transformation is approximately given by
\begin{align}
H_{(1,1)} \approx \begin{pmatrix}
\bar E_z & \frac{E_{2,\perp}^*}{2} & \frac{E_{1,\perp}^*}{2} & 0 \\
\frac{{E_{2,\perp}}}{2} & \frac{\Delta E_z}{2} - \alpha_+ & \frac{\alpha_+ + \alpha_-}{2} & \frac{E_{1,\perp}^*}{2} \\
\frac{{E_{1,\perp}}}{2} & \frac{\alpha_+ + \alpha_-}{2} & -\frac{\Delta E_z}{2} - \alpha_- & \frac{E_{2,\perp}^*}{2} \\
0 & \frac{{E_{1,\perp}}}{2} & \frac{{E_{2,\perp}}}{2} & - \bar E_z.
\end{pmatrix}
\end{align}
Transforming to the logical adiabatic basis $\{e^{i\phi_{\uparrow\uparrow}(t)}\ket{\uparrow\uparrow},e^{i\phi_+(t)}\ket{\psi_+(t)}, e^{i\phi_-(t)}\ket{\psi_-(t)}, e^{i\phi_{\downarrow\downarrow}(t)}\ket{\downarrow\downarrow}\}$ using $\tilde H_{(1,1)} = R^\dagger H_{(1,1)} R + i \hbar (\partial_t R^\dagger) R$, where $R$ is a unitary transformation matrix whose rows are given by the logical basis states and $\phi_i(t)$ are phases associated with the choice of logical basis (which correspond to shifts in $ZZ$, $IZ$, $ZI$ generators of unitary time evolution operator), we obtain the logical basis Hamiltonian as
\begin{widetext}
\begin{align}
\tilde H_{(1,1)} \approx \begin{pmatrix}
\bar E_z + \dot \phi_{\uparrow\uparrow} & \frac{(\tilde E_{2,\perp}^+)^*}{2} e^{-i(\phi_{\uparrow \uparrow} - \phi_+)} & \frac{(\tilde E_{1,\perp}^+)^*}{2} e^{-i(\phi_{\uparrow \uparrow} - \phi_-)} & 0 \\
\frac{\tilde E_{2,\perp}^+}{2} e^{i(\phi_{\uparrow \uparrow} - \phi_+)} &  \frac{1}{2}(-J + \Delta E) + \dot\phi_+ & \tilde V & \frac{(\tilde E_{1,\perp}^-)^*}{2} e^{i(\phi_{\downarrow \downarrow}-\phi_+)} \\
\frac{\tilde E_{1,\perp}^+}{2} e^{i(\phi_{\uparrow \uparrow} - \phi_-)} & \tilde V^* &  \frac{1}{2}(-J - \Delta E) + \dot\phi_-& \frac{(\tilde E_{2,\perp}^-)^*}{2}e^{i(\phi_{\downarrow \downarrow}-\phi_-)} \\
0 & \frac{\tilde E_{1,\perp}^-}{2} e^{-i(\phi_{\downarrow \downarrow}-\phi_+)} & \frac{\tilde E_{2,\perp}^-}{2} e^{-i(\phi_{\downarrow \downarrow}-\phi_-)} & - \bar E_z + \dot \phi_{\downarrow\downarrow},
\end{pmatrix}
\end{align}
\end{widetext}
where the diabatic correction $\tilde V$ (given by $\approx i \hbar e^{-i(\phi_+ - \phi_-) } \frac{h_z \dot J - \dot h_z J}{h_z^2}$ when $J \ll h_z$) vanishes unless $J$ or $h_z$ is varying in time. The transverse terms in the logical adiabatic basis are given by
$\tilde E_{1,\perp}^\pm = \tilde \Omega_1^\pm e^{i \omega t} = \frac{ (\Delta E + h_z)\Omega_1 \mp J \Omega_2}{ \sqrt{2\Delta E (\Delta E + h_z)}} e^{i \omega t}$ and
$\tilde E_{2,\perp}^\pm = \tilde \Omega_2^\pm e^{i \omega t} = \frac{(\Delta E + h_z) \Omega_2  \pm J \Omega_1 }{ \sqrt{2\Delta E (\Delta E + h_z)} } e^{i \omega t}$.
In the limit of $J \ll h_z$, they can be approximated as $\approx (\Omega_1  \mp \Omega_2 J/2 h_z) e^{i \omega t}$ and $(\Omega_2  \pm \Omega_1 J/2 h_z) e^{i \omega t}$ respectively.

One practical choice of logical frame is  $\dot\phi_\pm = -(-J_0 \pm \Delta E_0)/2 = -(-J \pm \Delta E)/2|_{\epsilon=\epsilon^0}$ and $\dot\phi_{\uparrow\uparrow} =\dot\phi_{\downarrow\downarrow}= -\bar E_z|_{\epsilon=\epsilon^0}$ such that when the ESR line is turned off, there would be no evolution at $\epsilon=\epsilon_0$. However, different choices are equally valid.
We remark that the logical frame itself, which is fixed once the choice is made, should not depend on the control for a general purpose quantum computer.

EDSR allows separate control over $\Omega_i$, which would allow a straightforward control over each qubit when $J/h_z$ is small enough. With ESR, however, this is not possible and when pulsing only the ESR current, the ratio $\Omega_2/\Omega_1$ is a fixed number close to 1. Furthermore, when $J/h_z$ is not small enough (e.g., when using an always-on exchange \cite*{Huang2018}, or when the residual minimal exchange is non-negligible), nonlocal terms $ZX,ZY,XZ,YZ$ in this Hamiltonian lead to crosstalk among qubits.
This problem can be addressed as follows.

Let us assume we would like to address the second qubit in order to implement the $IX$ rotation in the main text; the procedure for addressing the first qubit is basically the same, with the order of qubits swapped. To do that, we tune the microwave frequency to $\omega = \omega_0 + \delta \omega $ with $\hbar \omega_0 = (\bar E_z-\Delta E/2)|_{\epsilon=\epsilon^0}$ such that
\begin{widetext}
\begin{align}
\tilde H_{(1,1)} = \begin{pmatrix}
E_z' & \frac{\tilde\Omega_2^+}{2} e^{\frac{i}{\hbar} [J_0 /2 - \hbar\delta \omega ]t} & \frac{\tilde\Omega_1^+}{2} e^{\frac{i}{\hbar} (\Delta E_0 + J_0/2 - \hbar\delta \omega)t } & 0 \\
\frac{\tilde\Omega_2^+}{2} e^{-\frac{i}{\hbar}[J_0/2 - \hbar\delta \omega ]t} &  (-J' + \Delta E')/2 & 0 & \frac{\tilde\Omega_1^-}{2} e^{\frac{i}{\hbar}(\Delta E_0 - J_0/2 - \hbar\delta \omega)t} \\
\frac{\tilde\Omega_1^+}{2} e^{-\frac{i}{\hbar} (\Delta E_0 + J_0/2 - \hbar\delta \omega)t } & 0 &  (-J' - \Delta E')/2 & \frac{\tilde\Omega_2^-}{2}e^{\frac{i}{\hbar} [-J_0 t/2- \hbar\delta \omega ]t} \\
0 & \frac{\tilde\Omega_1^-}{2} e^{-\frac{i}{\hbar}(\Delta E_0 - J_0/2- \hbar\delta \omega)t} & \frac{\tilde\Omega_2^-}{2} e^{-\frac{i}{\hbar} [-J_0/2- \hbar\delta \omega]t} & -E_z'
\end{pmatrix},
\end{align}
\end{widetext}
where $J' = J - J_0$, $\Delta E' = \Delta E - \Delta E_0$ and $\bar E_z' = \bar E_z - \bar{ E_z}_0$. For simplicity, we will take $\Delta E' = \bar E_z' = 0$.
When $\Delta E_0 - \hbar\delta\omega \gg \Omega_1^\pm$, the $\Omega_1^\pm$ terms can be neglected as fast oscillating terms. When this is not the case, as for example in Ref.~\onlinecite*{Huang2018}, these terms would lead to systematic errors, and for the purpose of estimating these errors, we split the total Hamiltonian as $H_0 + H_I$ where the ``interaction Hamiltonian" $H_I$ contains the $ZI$ and $\Omega_1^\pm$ terms, and $H_0$ contains the remaining terms. The time-evolution operator can formally be written as $U = U_0 U_I$ where
\begin{align}
U_0 =  \mathcal T e^{-\frac{i}{\hbar} \int dt H_0}, \qquad U_I = \mathcal T e^{-\frac{i}{\hbar} \int dt U_0^\dagger H_I U_0 }.
\label{eq:interaction}
\end{align}
and $\mathcal T$ is the time-ordering operator. $U_I$ can be seen as the error propagator, and for small enough $\Omega_1^\pm/\Delta E$, one can use the lowest order Magnus expansion to evaluate it as $U_I = e^{-\frac{i}{\hbar} \int dt U_0^\dagger H_I(t) U_0 }$.

Generally speaking, the problem of calculating a time-evolution operator may be expressed in a relatively nicer looking form if in a different frame. In a frame rotated by $R$, the time-evolution operator and Hamiltonian become
\begin{align}
U_R = R^\dagger U \to U = R U_R, \quad H_R = R^\dagger H R + i (\partial_t R^\dagger) R.
\end{align}
If calculating $U_R$ is a simpler problem, we can calculate it first and obtain the time-evolution operator in the original frame  as $R U_R$.

For $U_0$, we use the intermediate frame $R_0 = e^{\frac{i}{\hbar} [ZZ J_0 /4 - IZ \hbar\delta \omega/2] t}$, which yields the rotated Hamiltonian
\begin{align}
\tilde H_0 =& \frac{\tilde\Omega_2^+ + \tilde\Omega_2^-}{4} IX + \frac{\tilde\Omega_2^+ - \tilde\Omega_2^-}{4} ZX + \nonumber \\
& \frac{J}{4} ZZ  - IZ \frac{\hbar(\delta \omega + \delta \dot{\omega} t)}{2}.
\end{align}
In what follows, we will take a constant frequency for simplicity, $\delta \dot {\omega}=0$, although we can also take $\delta \omega = \delta\omega_0 + \varphi_0/t$ to gradually shift the microwave frequency during the one-qubit gate operation. This Hamiltonian can be written using two distinct $\mathfrak{su}(2)$ algebras, so by rearranging terms that way we obtain
\begin{align}
U_0 =&  R_0 \mathcal T e^{-\frac{i}{\hbar} \int dt \tilde H_0} \nonumber\\
= &
e^{\frac{i}{\hbar} [ZZ J_0 /4 - IZ \frac{i}{\hbar} \delta \omega/2] t} \times \nonumber \\
& \left( \mathcal T e^{-\frac{i}{\hbar} \int dt \left[\frac{\tilde\Omega_2^+ + \tilde\Omega_2^-}{4} IX + \frac{\tilde\Omega_2^+ - \tilde\Omega_2^-}{4} ZX + \frac{J}{4} ZZ - IZ \frac{\hbar \delta \omega}{2} \right] } \right)  \nonumber\\
 =& e^{\frac{i}{\hbar} [ZZ J /4 - IZ \hbar \delta \omega/2] t} \times \nonumber\\
& \mathcal T e^{-\frac{i}{\hbar} \int dt \left[\frac{\tilde\Omega_2^+}{2}\frac{IX+ZX}{2}  +  \left(\frac{J}{4} - \frac{\hbar \delta \omega}{2} \right) \frac{IZ+ZZ}{2}\right]} \times \nonumber\\
& \mathcal T e^{-\frac{i}{\hbar} \int dt \left[\frac{\tilde\Omega_2^-}{2}\frac{IX-ZX}{2}  + \left(-\frac{J}{4} - \frac{\hbar\delta \omega}{2} \right) \frac{IZ-ZZ}{2}\right]}.
\label{eq:U0}
\end{align}
While the solution is straightforward for a square pulse on ESR power, there are also known solutions of the Bloch equation corresponding to the SU(2) Hamiltonian $H = f(t) \sigma_x + c \sigma_z$ \cite*{Bagrov2005,Barnes2012,Barnes2013,Barnes2015} for certain types of envelope functions $\Omega_2^\pm(t)$. Either way, the gate time and the envelope or pulse amplitude should be chosen in such a way that we target a specific $IX$ rotation angle and the crosstalk term $ZX$ vanish at the final gate time. This condition can be written by using Euler decomposition for each SU(2) time-evolution operator:
\begin{widetext}
\begin{align}
U_0 =& e^{\frac{i}{\hbar} [ZZ J /4 - IZ \delta \hbar\omega/2] t} \left(
e^{i \alpha_1^+ \frac{IZ+ZZ}{2}} e^{i \alpha_2^+ \frac{IX+ZX}{2}} e^{i \alpha_3^+ \frac{IZ+ZZ}{2}}
\right)
\left(
e^{i \alpha_1^- \frac{IZ-ZZ}{2}} e^{i \alpha_2^- \frac{IX-ZX}{2}} e^{i \alpha_3^- \frac{IZ-ZZ}{2}}
\right) \nonumber\\
=& e^{\frac{i}{\hbar} [ZZ J /4 - IZ \hbar\delta \omega/2] t} \left(
e^{i \left[\alpha_1^- \frac{IZ-ZZ}{2} + \alpha_1^+ \frac{IZ+ZZ}{2}\right]}   e^{i \left[\alpha_2^+ \frac{IX+ZX}{2} + \alpha_2^- \frac{IX-ZX}{2}\right]}
 e^{i \left[\alpha_3^- \frac{IZ-ZZ}{2} + \alpha_3^+ \frac{IZ+ZZ}{2}\right]}.
\right)
\end{align}
\end{widetext}
When $\Omega_2^\pm$ are time-independent, the angles $\alpha_i^\pm$ are given by
\begin{align}
\alpha_1^\pm = \alpha_3^\pm =& \frac{1}{2}\arctan(\cos\theta_\pm,\sin\theta_\pm \cos\phi_\pm), \nonumber\\
\alpha_2^\pm =& \arctan\left(\sqrt{1 - \sin^2\theta_\pm \sin^2\phi_\pm}, \sin\theta_\pm \sin\phi_\pm\right),
\end{align}
where
\begin{align}
\cos\phi_\pm =& -\frac{\pm J/4 - \hbar\delta\omega/2}{\hbar\omega_\pm},\qquad \sin\phi_\pm = -\frac{\tilde\Omega_2^\pm/2}{\hbar\omega_\pm}, \nonumber\\
\theta_\pm =&  \omega_\pm t, \qquad \hbar\omega_\pm = \sqrt{(\pm J/4 - \hbar\delta\omega/2)^2 + (\tilde\Omega_2^\pm/2)^2}
\end{align}
and $\arctan(x,y)$ is the two-parameter arc-tangent function.

When $\alpha_2^+ = \alpha_2^-$, the crosstalk is removed, making $U_0$ equivalent to an $IX$ rotation which is surrounded by $IZ$ and $ZZ$ from both sides.
Targeting a $\theta_0$ rotation around $IX$ without any $ZX$ rotations respectively correspond to the following constraints:
\begin{align}
\frac{\alpha_2^+ + \alpha_2^-}{2} = \theta_0, \qquad \frac{\alpha_2^+ - \alpha_2^-}{2} = 0.
\label{eq:crosstalk}
\end{align}
These constraints can be solved for $\Omega_2^\pm$, $\delta \omega$ and $t$. We note that since the ratio of $\Omega_1$ to $\Omega_2$ is fixed, $\Omega_2^\pm$ corresponds to a single degree of freedom, thus the solution contains a single free parameter, which can be taken to be $\delta \omega$ without any loss of generality, and used to target a specific $IZ$ rotation.

While the surrounding $IZ$  and $ZZ$ rotations can be canceled by using the exchange and the microwave frequency when needed, if an $IX$ gate is surrounded by $ZZ$ or $IZ$ rotations in a pulse sequence, they also can be used to reduce the execution times of the neighboring gates. For the purposes of our pulse sequence, the microwave frequency should be chosen in such a way that the $IZ$ rotations which accompany the $IX$ rotations cancel their neighboring $IZ$ rotation which accompany the middle $ZZ_{\pi/2}$ rotation. There are also $IZ$ rotations which neighbor the outer $ZZ$ rotations, but since $IZ$ commutes with $ZZ$, they can be taken outside.
This is similar to the ``virtual" one-qubit $Z$ gates \cite*{McKay2017,Knill2008,Vandersypen2005}: by shifting the microwave frequency, we can have additional $Z$ gates which surround the original gate at no cost.

In the presence of noise, a concatenated pulse sequence such as CinS \cite*{Bando2013} or a robust shaped pulse \cite*{Barnes2015} can be used to correct the $IZ$ and $ZZ$ errors in the Hamiltonian caused by charge noise and nuclear spins. However, $ZI$ errors cannot be fixed this way since $H_0$ commutes with $ZI$.

A detailed characterization of this gate will be provided in a subsequent work.

\section{Echo schemes to robustly remove unwanted local $Z$ rotations from $U_\text{ad}(T)$}
\label{sec:echo}
If we have access to high-fidelity one-qubit $\pi$-pulses, which can be realized by employing shaped pulses or pulse sequences, we can echo out the unwanted rotations using $\pi$-pulses
\begin{align}
U_{J,1}(T) =
& U_\text{ad}\left(\frac{T}{2}\right) e^{-i \frac{\pi}{2} IX} e^{-i \frac{\pi}{2} XI} U_\text{ad}\left(\frac{T}{2}\right) e^{-i \frac{\pi}{2} IX} e^{-i \frac{\pi}{2} XI} \nonumber\\
=& e^{-i \frac{\pi}{2} IX} e^{-i \frac{\pi}{2} XI} U_\text{ad}\left(\frac{T}{2}\right) e^{-i \frac{\pi}{2} IX} e^{-i \frac{\pi}{2} XI} U_\text{ad}\left(\frac{T}{2}\right) \nonumber\\
=& e^{-i \tilde\gamma(T) ZZ}.
\label{eq:pi-pulse1}
\end{align}
and obtain a pure $ZZ$ rotation, for arbitrary $\bar E_z, \Delta E_z$. Note that $\pi$-pulses cancel quasistatic errors in both  $ZI-IZ$ and $ZI+IZ$ terms to all orders. The end result for the local operations is a robust identity.

If, however, this pulse sequence is implemented using non-robust one-qubit gates, the average infidelity of a perfect entangler implemented using $U_{J,1}(T)$ comes out to be
\begin{align}
\langle 1-\mathcal F_\text{seq}\rangle \approx   \iint d\delta\Omega_1 d\delta\Omega_2 p(\delta\Omega_1,\delta\Omega_2) \times \nonumber\\
\left[ 23.94 \left(\frac{\delta \Omega_1}{\Omega_1}\right)^2 + 19.07 \left(\frac{\delta \Omega_2}{\Omega_2}\right)^2 + 2.64 \frac{\delta \Omega_1 \delta \Omega_2}{\Omega_1 \Omega_2} \right].
\end{align}
Assuming a multivariate Gaussian distribution for errors on Rabi frequencies with standard deviations $\sigma_{\delta \Omega_i}$ and covariance $\rho \sigma_{\Omega_1} \sigma_{\Omega_2}$ where $\rho \leq 1$, we obtain $\langle 1-\mathcal F_\text{seq} \rangle \approx 23.94 (\sigma_{\delta \Omega_1}/\Omega_1)^2 + 19.07 (\sigma_{\delta \Omega_2}/\Omega_2)^2 + 2.64 \rho \sigma_{\Omega_1} \sigma_{\Omega_2}/\Omega_1 \Omega_2$ for the infidelity of the perfect entangler gate.

It is also possible to implement this robust trivial local dynamics using non-robust one-qubit gates by making use of noisy local $\pi$-pulses in $Z$ and $X$, such that
their first order errors cancel each other \cite*{Kabytayev2014}, resulting in a noisy nonlocal and robust local identity gates:
\begin{align}
U_{J,2} =& e^{-i \frac{\pi}{2} IX} e^{-i \frac{\pi}{2} XI} U_{J,\pi,\gamma_1} e^{-i \frac{\pi}{2} IX} e^{-i \frac{\pi}{2} XI} U_{J,\pi,\gamma_2} \nonumber\\
=& e^{-i  (\gamma_1 + \gamma_2) ZZ},
\label{eq:pi-pulse2}
\end{align}
where $U_{J,\pi,\gamma_i} \equiv e^{-i \gamma_i ZZ} e^{-i  \frac{\pi}{2} ZI} e^{-i  \frac{\pi}{2} IZ}$
This can be relevant in the SiMOS setup \cite*{Veldhorst2014} where $g$-factors are electrically modulated, making all local rotations susceptible to charge noise.
In the presence of quasistatic noise, the average infidelity of this sequence is given by Eq.~\eqref{eq:fidelity}.

We remark that  $ZI$ rotations do not need to vanish since they commute with the pulse sequence and can effectively be moved out.

\section{Perturbative filter function for $\mathfrak{su}(N)$}
\label{sec:filter-function}
In this Appendix, we derive the leading order filter function $F^{(1)}$ for weak noise and short times for a $\mathfrak{su}(N)$ Hamiltonian. The presentation here is a generalized version of \cite*{Green2013}.

Given a control Hamiltonian $H_c$ and a noise Hamiltonian $H_\epsilon$,
\begin{align}
\tilde H=H_c + H_\epsilon, \quad H_c = \sum_i h_i \Lambda_i, H_\epsilon = \sum_i \epsilon_i \Lambda_i \equiv \sum_i \chi_i \beta_i \Lambda_i,
\end{align}
where $\beta_i$ is the stochastic part of $H_\epsilon$ and $\Lambda_i$ are $\mathfrak{su}(N)$ generators, the noisy time-evolution operator $\tilde U$ can be written by treating $H_\epsilon$ as the ``interaction Hamiltonian" as (cf. Eq.~(\ref{eq:interaction}))
\begin{align}
\tilde U = U_c U_\epsilon, \qquad U_\epsilon = \mathcal T e^{-\frac{i}{\hbar}\int_0^T  U_c^\dagger(t) H_\epsilon(t) U_c(t)}.
\end{align}
When the noise Hamiltonian and the total evolution time $T$ are small enough such that $U_\epsilon$ is sufficiently close enough to identity, one can use the first order Magnus expansion to evaluate $U_\epsilon$ as $\approx e^{-\frac{i}{\hbar} \int_0^T U_c^\dagger(t) H_\epsilon(t) U_c(t)}$ (further information regarding the convergence of the Magnus expansion can be found in Ref.~\onlinecite*{Green2013}). The average leading order trace fidelity can then be written as
\begin{align}
\langle \mathcal F_\text{tr} \rangle = \frac{\langle \text{tr}(U_c \tilde U) \rangle}{\text{tr}{(\Lambda_i \Lambda_i)}} = 1 - \langle a_1^2 \rangle + \ldots
\end{align}
where
\begin{align}
\langle a_1^2 \rangle = \langle\text{tr}\int_0^T \int_0^T dt_1 dt_2
&\left[U_c^\dagger(t_1) \frac{H_\epsilon(t_1)}{\hbar} U_c(t_1)\right] \nonumber \\
&\left[U_c^\dagger(t_2) \frac{H_\epsilon(t_2)}{\hbar} U_c(t_2)\right]\rangle.
\end{align}
Using the $N^2-1$ dimensional adjoint representation of $U_c^\dagger$ defined through $(R \boldsymbol \epsilon) \cdot \boldsymbol \Lambda \equiv U_c^\dagger (\boldsymbol \epsilon \cdot \boldsymbol \Lambda) U_c = H_\epsilon^{(I)}$, or alternatively
\begin{align}
R_{ij}(t) = \frac{\text{tr}(\Lambda_i U_c^\dagger(t) \Lambda_j U_c(t))}{\text{tr}(\Lambda_i \Lambda_i)},
\end{align}
gate infidelity can be compactly rewritten as
\begin{align}
\langle a_1^2 \rangle =& \left\langle \int_0^T \int_0^T dt_1 dt_2  \left[ R(t_1) \frac{\boldsymbol \epsilon(t_1)}{\hbar} \right] \cdot \left[ R(t_2) \frac{\boldsymbol \epsilon(t_2)}{\hbar} \right] \right\rangle \nonumber\\
=& \frac{1}{\hbar^2} \sum_{i,j,k} \int_0^T \int_0^T dt_1 dt_2
\langle \beta_i(t_1) \beta_j(t_2) \rangle  \times\nonumber\\
& \qquad \qquad \qquad \chi_i(t) \chi_j(t) R_{ki}(t_1) R_{kj}(t_2).
\end{align}
In the frequency domain, gate infidelity can be expressed in terms of the PSD $S_{ij}(\omega)$ defined through $\langle \beta_i(t_1) \beta(t_2)\rangle = \frac{1}{2\pi}\int_{-\infty}^\infty d\omega S_{ij} e^{i\omega (t_2-t_1)}$ (assuming the autocorrelation function only depends on the difference $t_2 - t_1$), and using this relation to replace the stochastic terms with the PSD, and defining the frequency-domain ``control matrix" $R_{ki}(\omega) \equiv -i\omega \int_0^T dt R_{ki}(t) \chi_i(t) e^{i\omega t}$, we finally reach to the following expression for the gate infidelity:
\begin{align}
\langle 1-\mathcal F_\text{tr} \rangle \approx \frac{1}{\hbar^2}\sum_{i,j,k}\int_{-\infty}^{\infty} \frac{d\omega}{2\pi} S_{ij}(\omega) \frac{R_{kj}(\omega) R^*_{ki}(\omega)}{\omega^2}.
\end{align}
We identify the $\sum_k R_{kj}(\omega) R_{ki}^*(\omega) = [R^\dagger(\omega) R(\omega)]_{ij}$ term as the first-order filter-function $F^{(1)}_{ij}(\omega)$. Higher order corrections to the infidelity involving higher order filter functions can be obtained in a similar fashion as described in Ref.~\onlinecite*{Green2013}.

We remark that the adjoint representation $R$ can be block diagonalized when $H_c$ belongs to a subalgebra of $\mathfrak{su}(N)$ \cite{Gungordu2012a}.

\bibliography{siqd,extra}

\end{document}